\documentclass[preprint,12pt,authoryear]{elsarticle}

\usepackage{graphicx}
\usepackage{caption}
\usepackage{subcaption}
\usepackage{amsmath,amsfonts}
\usepackage{array}
\usepackage{textcomp}
\usepackage{stfloats}
\usepackage{url}
\usepackage{verbatim}
\usepackage{balance}
\usepackage{lineno} 
\usepackage{array}
\usepackage{booktabs}
\usepackage{tikz}
\usetikzlibrary{positioning, fit, shapes.geometric, arrows}
\usepackage{colortbl}       

\journal{Machine Learning with Applications} 

\begin{document}

\begin{frontmatter} 

\title{ Deep Learning-Based Identification of Precipitation Clouds from All-Sky Camera Data for Observatory Safety}

\author[label1,label2]{Mohammad H. Zhoolideh Haghighi\corref{cor1}}
\ead{zhoolideh@kntu.ac.ir}
\cortext[cor1]{Corresponding author}
\address[label1]{Department of Physics, K.N. Toosi University of Technology, Tehran P.O. Box 15875-4416, Iran}
\address[label2]{School of Astronomy, Institute for Research in Fundamental Sciences (IPM), Tehran, 19395–5746, Iran}

\author[label1]{Alireza Ghasrimanesh}\ead{alireza.ghasrimanesh@email.kntu.ac.ir}
\author[label2]{Habib Khosroshahi}\ead{habib@ipm.ir}

\begin{abstract}
	For monitoring the night sky conditions, wide-angle all-sky cameras are used in most astronomical observatories to monitor the sky cloudiness. In this manuscript, we apply a deep-learning approach for automating the identification of precipitation clouds in all-sky camera data as a cloud warning system.  We construct our original training and test sets using the all-sky camera image archive of the Iranian National Observatory (INO). The training and test set images are labeled manually based on their potential rainfall and their distribution in the sky. We train our model on a set of roughly 2445 images taken by the INO all-sky camera through the deep learning method based on the EfficientNet network. Our model reaches an average accuracy of 99\% in determining the cloud rainfall's potential and an accuracy of 96\% for cloud coverage. To enable a comprehensive comparison and evaluate the performance of alternative architectures for the task, we additionally trained three models—LeNet, DeiT, and AlexNet. This approach can be used for early warning of incoming dangerous clouds toward telescopes and harnesses the power of deep learning to automatically analyze vast amounts of all-sky camera data and accurately identify precipitation clouds formations. Our trained model can be deployed for real-time analysis, enabling the rapid identification of potential threats, and offering a scaleable solution that can improve our ability to safeguard telescopes and instruments in observatories. This is important now that numerous small- and medium-sized telescopes are increasingly integrated with smart control systems to reduce manual operation.
\end{abstract}


\end{frontmatter}


\balance 

\section{Introduction}
The field of astronomy relies heavily on advanced telescopes and astronomical instruments to observe celestial phenomena and unravel the mysteries of the cosmos. However, these sophisticated devices, particularly ground-based telescopes, are susceptible to the adverse effects of hazardous atmospheric conditions, such as rainfall, snowfall, and humidity, which can harm their optical surfaces and electronic components. Therefore, timely identification of such natural occurrences is crucial to safeguard the integrity and longevity of these valuable instruments. Clouds, of course, can prevent optical telescopes from carrying out observations. This, in turn, can impede observations and compromise the quality of data collected.

Cloud coverage is a key factor impacting astronomical observations from the ground in optical and infrared regimes, as cloudy conditions significantly reduce the time during which a telescope can be effectively utilized.

Telescopes are also costly instruments that require protection from environmental factors, making it essential to promptly close the telescope dome during adverse weather while ensuring it remains open during partially cloudy skies to avoid missed observation opportunities. To address these challenges, in addition to using satellite meteorological data, many prominent astronomical observatories utilize all-sky cameras to monitor sky conditions and assess cloud coverage distribution. These devices use affordable charge-coupled devices (CCDs) and wide-angle lenses for regular monitoring.

Traditional methods for monitoring and detecting precipitation cloud formations around astronomical observatories typically involve either manual observation or basic automated systems. However, these approaches may fall short in terms of the speed and precision required for timely preventive actions, often relying heavily on human judgment. As numerous small- and medium-sized telescopes are increasingly integrated with automated or smart control systems, a robust and reliable automated cloud monitoring system has become essential.

When it comes to analyzing all-sky camera images, various methods have been developed to assess cloud coverage. For example, pixel segmentation approaches evaluate cloud distribution by analyzing intensity differences between cloudy and clear sky regions. \cite{afiq2019} categorized nighttime cloud images into clear and cloudy regions using the peak value of the all-sky image histogram.  \cite{jadhav2015} implemented Gaussian fitting and thresholding methods for binary segmentation of cloud images.  \cite{dev2017} applied superpixel techniques for binary segmentation of nighttime images. 
Fortunately, with the rapid advancements in computing power and pattern recognition technology, machine learning-based automatic cloud image recognition has become a prominent research area. For instance, a novel deep convolutional neural network, CloudU-Net \citep{shi2021}, has been introduced for binary segmentation of cloud images. \cite{li2022} applied machine learning models such as support vector machines, K-nearest neighbors, decision trees, and random forests for automated cloud classification, using features like cloud weight, cloud area ratio, and cloud dispersion. Additionally, two machine learning models, Light Gradient Boosting Machine (LightGBM) and ResNet, have been employed for binary cloud image classification \citep{momert}. 
As demonstrated by studies such as \cite{Guzel} and \cite{Yousaf}, the application of deep learning in atmospheric science has led to improved accuracy in cloud classification and severe weather pattern identification. Also, \cite{Glennen} presents an approach to cloud classification for air traffic control, and  \cite{calbo} classifies night sky conditions based on cloud features.

 Applying machine and deep learning techniques to all-sky camera images offers a proactive solution for identifying and predicting the presence and distribution of precipitation clouds, which is crucial for automating the opening and closure of telescope domes.
A key advantage of automating cloud identification is that, depending on the clouds' position and type, observations can continue in specific areas of the sky even during cloudy nights. Moreover, by monitoring the number of clear nights and cloud coverage distribution, the total amount of observable time throughout the year can be accurately estimated.

This manuscript aims to present a fully automated algorithm capable of classifying all-sky images in real-time with high accuracy. We employ advanced deep-learning techniques to identify cloud types based on their precipitation potential and spatial distribution within all-sky camera images. Through continuous analysis of data from all-sky cameras, the model can detect precipitation clouds, enabling observatories to implement preventive actions, such as pausing observations or activating protective protocols, to minimize potential damage or degradation of data quality. For training models, we utilize deep learning approach based on the EfficientNet \citep{efficientnet}, LeNet-5 \citep{lecun1998gradient}, AlexNet \citep{krizhevsky2012imagenet} , and DeiT \cite[]{touvron2021training} architectures. 

The contributions of this paper are summarised as follows: In Section \ref{data}, we present an overview of the dataset used in the study. Section \ref{model} describes the architectures and specifics of the deep learning models applied in training. Finally, Section \ref{result} presents and analyses the results generated by the models.

\section{Data} \label{data}

\subsection{Cammera}
\begin{figure}
\centering
\includegraphics[width=6cm]{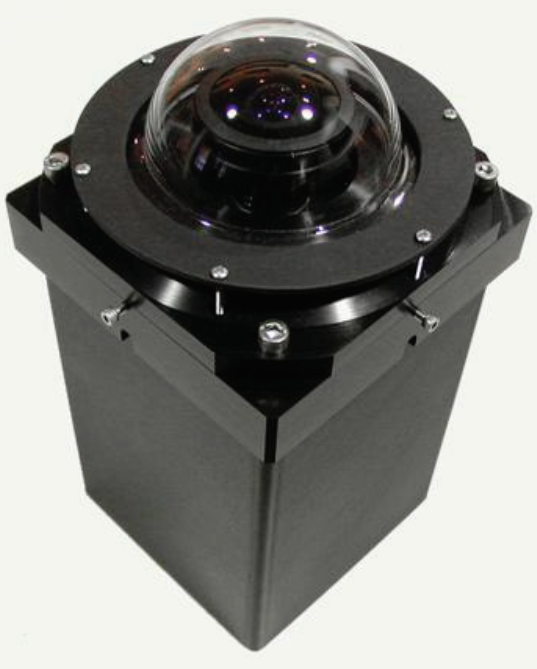}

\caption{All-sky 340 Camera.}
\label{All_Sky_340}
\end{figure}

We train and test our machine learning models using our own image data collected at the Iranian National Observatory (INO) site. The device we employ is an SBIG AllSky-340 CCD camera, equipped with a 640$\times $480 pixel CCD detector and a 1.4mm F/1.4 fish-eye lens, providing a 180° field of view. This camera produces a projection of the sky plane with 640$\times $480 pixels, each 7.4 microns square. The images are captured in JPG format with an exposure time of 60 seconds, and the imaging cadence at night is one frame per minute. Figure \ref{All_Sky_340} shows an AllSky-340 CCD camera. Iranian National Observatory is located at Mt Gargash 3600m above sea level and operates a 3.4m optical telescope \citep{Khosroshahi} and other astronomical facilities.

\subsection{Cloud Classes}

Cloud classification plays a vital role in understanding weather patterns and predicting precipitation. Clouds are generally categorized based on their altitude and appearance, with each class offering different insights into the likelihood of precipitation. The World Meteorological Organization (WMO) has classified clouds into ten major types, further divided into three main altitude groups: high clouds, middle clouds, and low clouds.
High clouds, such as cirrus, cirrostratus, and cirrocumulus, form above 6,000 meters. These clouds are composed primarily of ice crystals due to the colder temperatures at high altitudes. Cirrus clouds are thin, wispy clouds that generally do not produce precipitation on their own, although they can signal an approaching weather front. Cirrostratus clouds may cover the sky with a transparent veil and can occasionally be associated with light precipitation, particularly if they thicken ahead of an incoming frontal system \citep{WMO}.
Middle clouds, which include altostratus and altocumulus, form between 2,000 and 6,000 meters. Altostratus clouds often appear as a gray or blue-gray layer, covering the sky and usually preceding storms. They are associated with moderate precipitation, particularly rain or snow, if the cloud layer thickens \citep{stull2015}. Altocumulus clouds, while less likely to produce precipitation, can sometimes indicate instability in the atmosphere and the potential for localized showers or thunderstorms, especially in warm, humid conditions \citep{rogers2000}.
Low clouds, which form below 2,000 meters, include stratus, stratocumulus, and nimbostratus clouds. Stratus clouds are uniform, gray clouds that often cover the entire sky like a blanket. These clouds are typically associated with light drizzle or mist. Stratocumulus clouds, although primarily non-precipitating, may produce light rain or drizzle under certain conditions. Nimbostratus clouds, on the other hand, are thick, dark clouds that almost always bring continuous and widespread precipitation, including rain or snow, depending on the temperature \citep{wallace2006}.
In addition to these cloud types, there are clouds with significant vertical development, such as cumulus and cumulonimbus clouds. Cumulus clouds are generally fair-weather clouds, but when they grow larger in unstable atmospheric conditions, they can develop into cumulonimbus clouds. Cumulonimbus clouds are towering, multi-layered clouds often associated with thunderstorms and severe weather. These clouds can produce heavy rainfall, lightning, hail, and even tornadoes, making them one of the most critical cloud types in terms of precipitation potential \citep{markowski2010}.

Understanding these cloud classes is essential for meteorology and weather forecasting. However, our focus is on determining whether observed clouds by an all-sky camera lead to precipitation or not. So, rather than categorizing clouds into one of the ten standard types, we classify them based on their potential for rainfall into three categories: High\_Potential\_Fall, Low\_Potential\_Fall, and Clear\_Sky. To label our dataset, we select a time interval of approximately two hours before the onset of rain or snowfall. Clouds that result in precipitation are labeled as High\_Potential\_Fall, while those that do not are labeled as Low\_Potential\_Fall. Images without clouds are classified as Clear\_Sky.

In addition to classifying clouds based on their precipitation potential, we also categorize them according to their distribution in the sky. Following the approach of  \cite{li2022}, we define three cloud distribution types: Inner, Outer, and Covered, and label the INO all-sky images accordingly. For this purpose, we draw two circles on the images with zenith angles of 44.7° and 65°, as illustrated in Figure \ref{cloud_dist}. If the cloud cover within both the inner and outer circles is less than 50\%, we label the image as Inner. If the inner circle is mostly clear but clouds are present within the outer circle, we label it as Outer. If more than 50\% of the area within both circles is covered by clouds, we label it as Covered.

It is important to note that our analysis focuses on night images, as the telescope dome remains closed during the day, eliminating the need to train a model on images taken before sunset. Additionally, aside from cropping the images to a square 480$\times $ 480-pixel size, no other image pre-processing is applied. This is an advantage of our approach, as the captured images can be used for real-time analysis.

\subsection{Train/Test Data Sample}

\begin{figure}
\centering
\includegraphics[width=12cm]{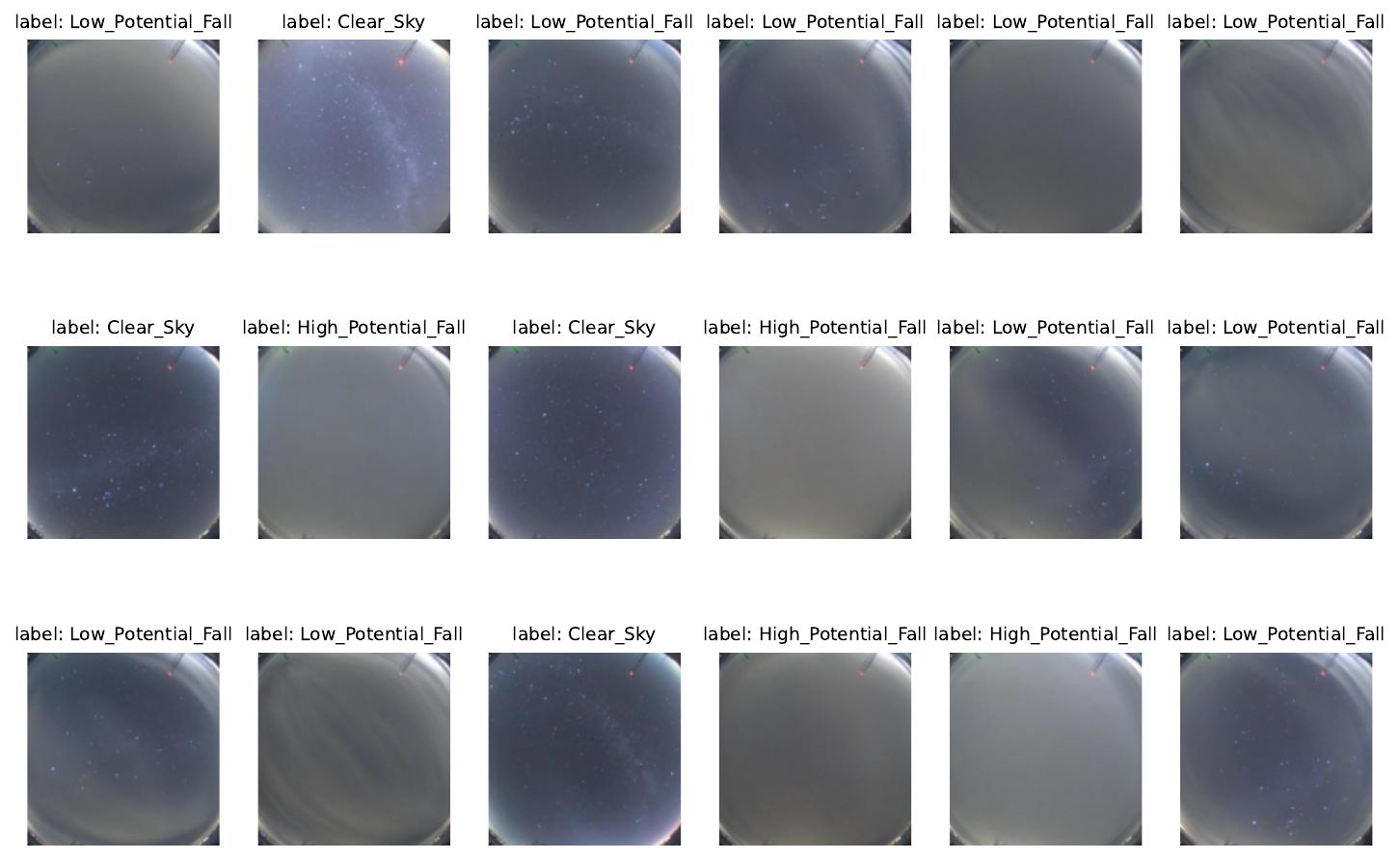}
\caption{Some of our training images with their corresponding labels.}
\label{cloud_fall}
\end{figure}

\begin{figure}
  \centering

  \captionsetup[subfigure]{labelformat=empty}  

  \begin{subfigure}{0.35\textwidth}
    \includegraphics[width=\linewidth]{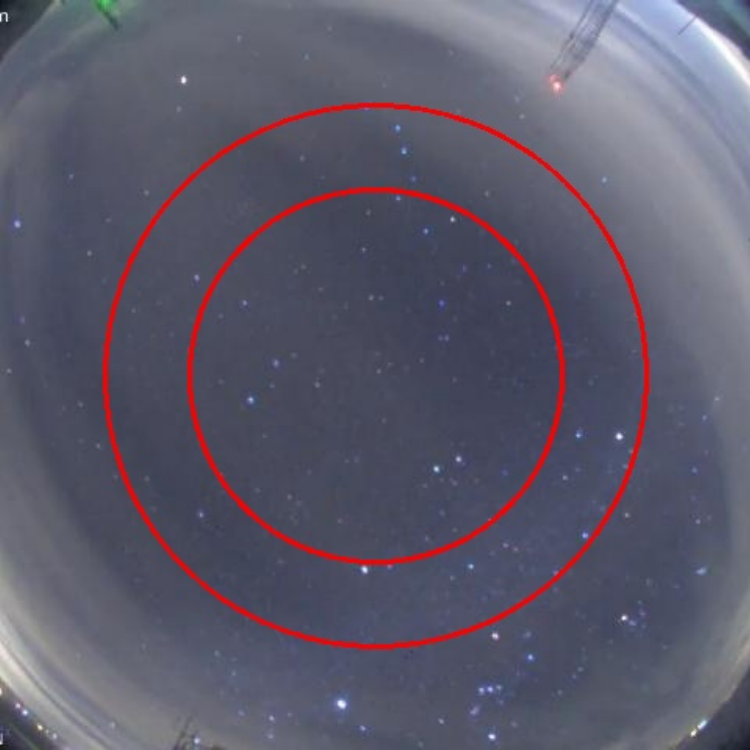}
    \caption{Inner}
    \label{fig:inner}
  \end{subfigure}
  \hfill
  \begin{subfigure}{0.35\textwidth}
    \includegraphics[width=\linewidth]{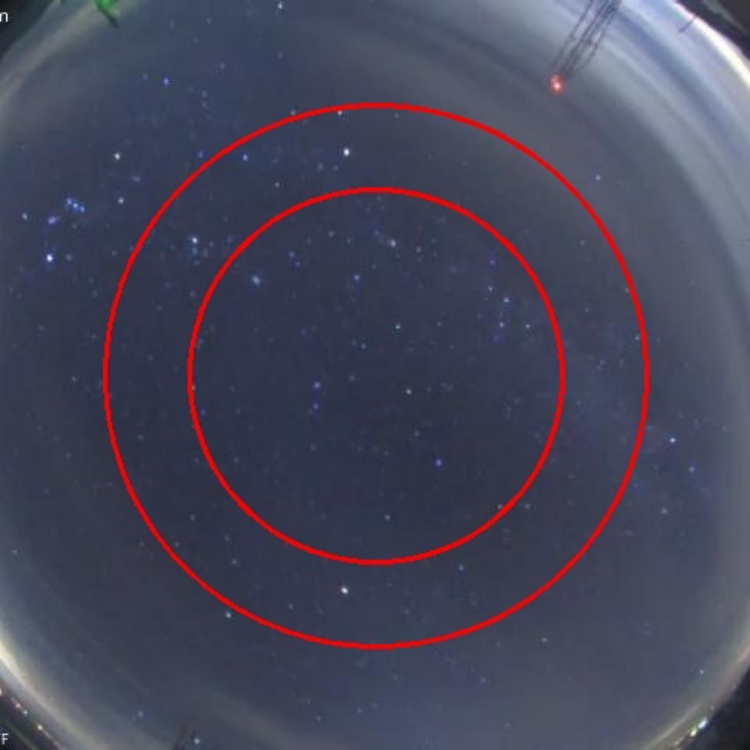}
    \caption{Outer}
    \label{fig:outer}
  \end{subfigure}
  \hfill
  \begin{subfigure}{0.35\textwidth}
    \includegraphics[width=\linewidth]{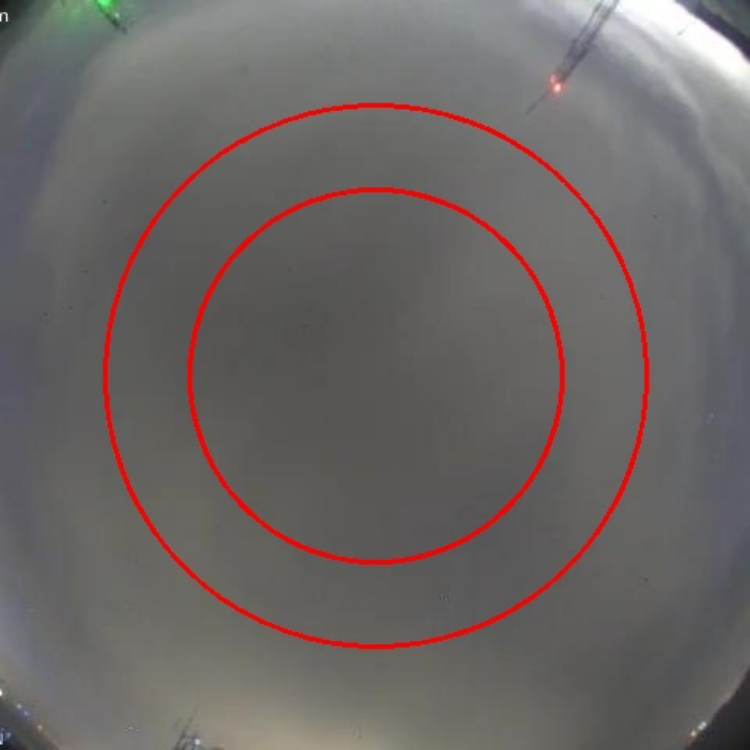}
    \caption{Covered}
    \label{fig:covered}
  \end{subfigure}

  \caption{Sample cloud distribution overlaid with zenith angle circles at 44.7° and 65°.}
  \label{cloud_dist}

\end{figure}

We selected 2,445 images for training, 482 for validation, and 516 for testing, all drawn from a dataset of images taken between January 2018 and December 2019. Two separate training sets were created: one for determining precipitation potential and another for cloud distribution. For the first set, we organized 2,030 manually labeled images into three classes: High\_Potential\_Fall, Low\_Potential\_Fall, and Clear\_Sky. For the second set, 1,413 images were labeled and grouped into three classes: Inner, Outer, and Covered. 
 Figures  \ref{cloud_fall}    \& \ref{cloud_dist} show examples of labeled images.

\section{Models definition \& training} \label{model}

\subsection{Models Architecture}

Neural networks, inspired by the structure and functioning of the human brain, form the backbone of modern machine learning algorithms. At their core, neural networks consist of interconnected nodes, or neurons, organized into layers. These layers typically include an input layer, one or more hidden layers, and an output layer. Information is processed through the network by assigning weights to the connections between neurons, and each neuron applies an activation function to its input, introducing non-linearity and complexity to the model.

The architecture of a neural network is a crucial determinant of its capabilities. Simple feedforward neural networks, where information flows in one direction from input to output, are fundamental building blocks. However, for more intricate tasks, deep neural networks, characterized by the inclusion of multiple hidden layers, have demonstrated unparalleled success. 
Training a neural network involves the optimization of its parameters, such as weights and biases, to minimize the difference between predicted outputs and actual target values. This process, known as backpropagation, entails iteratively adjusting the model's parameters based on the computed error. The choice of an appropriate loss function guides the optimization process by quantifying the difference between predictions and true values. Stochastic Gradient Descent (SGD) and its variants are popular optimization algorithms used to fine-tune the model efficiently \citep{Goodfellow}. 

In recent years, advancements in training methodologies have been propelled by the introduction of transfer learning. Transfer learning allows a pre-trained neural network on one task to be adapted for another, leveraging knowledge gained from one domain to boost performance in a related domain \citep{Yosinski}.
Instead of training a network from scratch and spending vast computing and time resources to find the best model, one can take advantage of the transfer learning approach. 
The fundamental idea behind transfer learning is to pre-train a neural network on a source task and then fine-tune it on a target task, utilizing the learned features and representations from the source task to enhance performance in the target domain. This approach is particularly beneficial when labeled data in the target domain is limited, as it allows the model to benefit from the abundance of data available in the source domain.
By leveraging pre-trained models, transfer learning significantly reduces the need for extensive labeled datasets in the target domain, mitigating challenges associated with data scarcity. 

In our approach, we focus on the EfficientNet neural network architecture \citep{efficientnet} as our ideal and preferred model, serving as the foundation for our methodology.
EfficientNets are state-of-the-art architectures commonly used in computer vision applications, including object identification, localization, classification, and image segmentation. Similar to all artificial neural networks, EfficientNet consists of several layers of neurons that are connected with each other and react to external stimuli. The network architecture of EfficientNet includes several convolutional layers, and the outputs of neurons affect both the neurons in the following layer and those in later layers.
It is based on the concept of compound scaling, which aims to balance the depth, width, and resolution of neural networks systematically. Previous approaches to scaling models either increased one of these dimensions—usually depth or width—independently, which often led to inefficiencies in both training and inference times. EfficientNet's compound scaling method overcomes this by scaling all three dimensions simultaneously using a simple but effective formula. This approach has led to substantial improvements in both accuracy and efficiency over traditional convolutional neural networks (CNNs).

The core idea behind EfficientNet is to use a compound coefficient to uniformly scale the network depth, width, and resolution. This allows the model to achieve better performance without drastically increasing computational complexity. The scaling rules are defined by three parameters: $\alpha$ for depth, $\beta$ for width, and $\gamma$ for resolution. The scaling factor $\phi$ controls how much each dimension is scaled, while the constants $\alpha$, $\beta$, and $\gamma$ ensure that the scaling is balanced across all dimensions. By employing this method, EfficientNet can create a family of models, ranging from EfficientNet-B0, the smallest and most efficient version, to EfficientNet-B7, which achieves state-of-the-art results on many benchmarks while maintaining reasonable computational demands \citep{efficientnet}.
 Additionally, EfficientNet includes squeeze-and-excitation (SE) blocks, which allow the network to focus on the most relevant features by recalibrating channel-wise feature responses. SE blocks have been shown to improve model accuracy while adding only a minimal amount of overhead \citep{hu2018squeeze}.
Another notable feature of EfficientNet is its use of the Swish activation function, instead of the more commonly used ReLU. The Swish function, defined as $f(x) = x \cdot \text{sigmoid}(x)$, enables smoother gradient flow during training, which can lead to better model convergence and higher accuracy, particularly in deep networks. Studies have demonstrated that Swish consistently outperforms ReLU in a variety of tasks, contributing to the overall success of the EfficientNet family \citep{ramachandran2017swish}.

Moreover, the scalability of EfficientNet has made it applicable across a wide range of domains. In medical imaging, for instance, it has been successfully applied to tasks such as cancer detection and retinal disease classification. These tasks require high accuracy while operating under constraints of limited computational power, making EfficientNet a perfect fit. Similarly, EfficientNet has been used in autonomous systems, such as drones and self-driving vehicles, where real-time decision-making requires both speed and precision \citep{chen2020efficientnet}.

We use the EfficientNet implementation from the Torchvision package, modifying the fully connected layer to match our target of three classes. The proposed network contains 63,802,326 parameters, as shown in Table \ref{tab:efficientnet_b7_classifier}, and its architecture is illustrated in Figure \ref{fig:efficientnet_architecture}.

\begin{table}[htbp]
	\centering
\caption{Layers and the corresponding parameters of the model.}
\label{tab:efficientnet_b7_classifier}
	\resizebox{\textwidth}{!}{%
		\begin{tabular}{|l|l|c|}
			\hline
			\textbf{Layer (Type)}              & \textbf{Output Shape} & \textbf{Number of Parameters} \\ \hline
			Input Layer                        & (480, 480, 3)         & 0                             \\ \hline
			EfficientNet-B7 (Base Model)       & (15, 15, 2560)        & 63,794,643                    \\ \hline
			Global Average Pooling Layer       & (2560)                & 0                             \\ \hline
			Dropout (Rate: 0.5)                & (2560)                & 0                             \\ \hline
			Dense (Fully Connected, 3 Units)   & (3)                   & 7,683                         \\ \hline
			\textbf{Total Parameters}          & \textbf{-}            & \textbf{63,802,326}           \\ \hline
		\end{tabular}%
	
	}

\end{table}

\usetikzlibrary{positioning, fit, shapes.geometric, arrows}

\tikzstyle{input} = [rectangle, rounded corners, minimum width=3cm, minimum height=1cm, text centered, draw=black, fill=blue!30]
\tikzstyle{block} = [rectangle, minimum width=3cm, minimum height=1cm, text centered, draw=black, fill=green!30]
\tikzstyle{subblock} = [rectangle, minimum width=2.5cm, minimum height=0.8cm, text centered, draw=black, fill=green!15]
\tikzstyle{dropout} = [rectangle, minimum width=3cm, minimum height=1cm, text centered, draw=black, fill=orange!30]
\tikzstyle{dense} = [rectangle, minimum width=3cm, minimum height=1cm, text centered, draw=black, fill=purple!30]
\tikzstyle{arrow} = [thick,->,>=stealth]

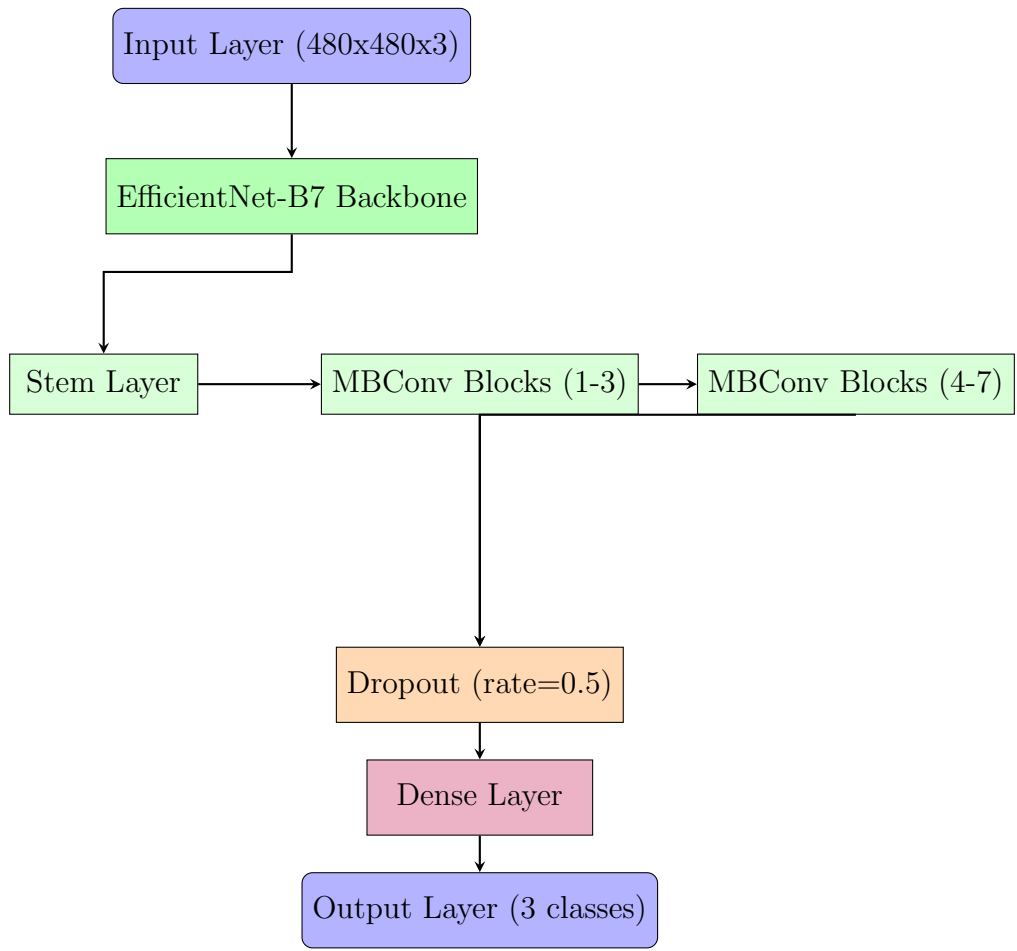
\begin{figure}

	\begin{tikzpicture}[node distance=1.5cm]
		
		\node (input) [input] {Input Layer (480x480x3)};
		
		\node (efficientnet) [block, below of=input, yshift=-0.5cm] {EfficientNet-B7 Backbone};
		
		\node (stem) [subblock, below of=efficientnet, xshift=-2.5cm, yshift=-1cm] {Stem Layer};
		\node (blocks1) [subblock, right of=stem, xshift=3.5cm] {MBConv Blocks (1-3)};
		\node (blocks2) [subblock, right of=blocks1, xshift=3.5cm] {MBConv Blocks (4-7)};
		
		\node (dropout) [dropout, below of=blocks1, yshift=-2.5cm] {Dropout (rate=0.5)};
		
		\node (dense) [dense, below of=dropout] {Dense Layer};
		
		\node (output) [input, below of=dense] {Output Layer (3 classes)};
		
		\draw [arrow] (input) -- (efficientnet);
		\draw [arrow] (efficientnet.south) -- ++(0,-0.5) -| (stem.north);
		\draw [arrow] (stem.east) -- (blocks1.west);
		\draw [arrow] (blocks1.east) -- (blocks2.west);
		\draw [arrow] (blocks1.south) -- (dropout.north);
		\draw [arrow] (blocks2.south) -| (dropout.north);
		\draw [arrow] (dropout) -- (dense);
		\draw [arrow] (dense) -- (output);
		
	\end{tikzpicture}
	
		\caption{Architecture of the EfficientNet-B7-based Model with Dropout and Dense Layers.}
\label{fig:efficientnet_architecture}
\end{figure}

Initially, the model was adapted to our specific problem by modifying the final fully connected layer to accommodate three output classes,  representing the High\_potential\_fall, Low\_potential\_fall, and Clear\_sky classes. We utilized mixed precision training with the \texttt{torch.cuda.amp} module to enhance computational efficiency and memory usage. The model was trained using the Adam optimizer \citep{Kingma2014Adam} with a  base learning rate of 0.001, and the loss function was set to \texttt{CrossEntropyLoss}. The training was conducted over 24 epochs, during which we monitored both training and validation losses, as well as accuracies.
 To prevent overfitting, dropout was applied in the final layer. The model's performance was evaluated using the $F_1$ score and ROC curves, with models being saved based on the best validation loss and $F_1$ score. The best model was selected based on minimum validation loss and highest score, demonstrating effective generalization to the validation dataset. Visualization of training, validation and test losses and accuracies, alongside the ROC curve, further highlighted the model's performance and robustness.

When it comes to the second model, we modified the final fully connected layer to output three classes including Inner, Outer, and Covered. Like what have done for the first model, we employed mixed precision training using the \texttt{torch.cuda.amp} module. 
The model was trained over 24 epochs using the Adam optimizer with a learning rate of 0.001. The loss function used was \texttt{CrossEntropyLoss}, and dropout was applied in the final layer. Training, validation and test losses, as well as accuracies, were recorded throughout the training process. 
The best models were saved based on the lowest validation loss and highest $F_1$ score. 

For the training of the machine learning models, we utilized a Tesla T4 GPU provided by Google Colab. This cloud-based platform offered efficient handling of large-scale computations and enabled the use of mixed precision training to optimize both computational speed and memory usage. The Tesla T4’s high performance significantly reduced training time, allowing for comprehensive experimentation and fine-tuning of the models.

While EfficientNet serves as our primary and ideal model due to its optimized scaling and high performance, it is essential to compare its effectiveness against other well-established architectures. To better understand the strengths and limitations of EfficientNet, we also evaluate the performance of LeNet-5, AlexNet, and DeiT. These models represent different stages in the evolution of deep learning, from early convolutional networks to modern transformer-based architectures. By comparing their training results, we aim to highlight the advantages of EfficientNet while analyzing how different architectural choices impact model performance.

LeNet-5 \citep{lecun1998gradient} is one of the earliest convolutional neural networks (CNNs), proposed by Yann LeCun and colleagues in 1998 for handwritten digit recognition, particularly on the MNIST dataset. The architecture consists of seven layers (excluding input), including convolutional, pooling (subsampling), and fully connected layers. LeNet-5 introduced the concept of local receptive fields, where neurons in a convolutional layer are only connected to a small region of the input, rather than being fully connected. This approach reduced the number of trainable parameters, making the network more computationally efficient.
A key innovation in LeNet-5 was the use of shared weights in convolutional layers, significantly reducing the number of parameters and improving generalization. Additionally, the model employed the sigmoid and hyperbolic tangent (tanh) activation functions to introduce non-linearity.

AlexNet's architecture \citep{krizhevsky2012imagenet}  is deeper and more complex than LeNet-5, containing eight layers: five convolutional layers followed by three fully connected layers. Unlike LeNet-5, which used sigmoid or tanh activations, AlexNet introduced the rectified linear unit (ReLU) activation function, which accelerated training by mitigating the vanishing gradient problem.
AlexNet addressed overfitting by incorporating dropout in the fully connected layers, randomly disabling neurons during training to force redundancy and improve generalization. Another innovation was the use of overlapping max pooling instead of traditional subsampling, which helped retain spatial information.

Transformers, originally introduced for natural language processing (NLP), have recently shown remarkable performance in computer vision. The Vision Transformer (ViT) \citep{dosovitskiy2020image} was one of the first models to apply transformer-based architectures to image classification, treating image patches as sequences and leveraging self-attention mechanisms. However, ViT required large datasets for training, limiting its applicability in scenarios with limited labeled data.
To address this limitation, Touvron et al. proposed the Data-Efficient Image Transformer (DeiT) \cite[]{touvron2021training}, which demonstrated that transformers can be trained efficiently on standard-sized datasets without requiring massive pretraining on large-scale datasets. DeiT introduced a novel distillation mechanism where a CNN-based teacher model guides the training of the transformer using a distillation token. This additional token interacts with the self-attention mechanism and helps the model learn more effectively from the teacher network.
Compared to CNNs, DeiT maintains competitive accuracy while benefiting from the inherent advantages of transformers, such as long-range dependencies and global attention. Additionally, it employs strong data augmentation techniques, including RandAugment and stochastic depth, further improving its robustness.

\subsection{Model Evaluation}

\subsubsection{Precision, Recall and $F_{1} Score$}

In classification, the predictions can result in distinct outcomes: \textit{True Positives (TP)}, where observations are correctly predicted as belonging to the target class; \textit{True Negatives (TN)}, where observations are correctly predicted as not belonging to the target class; \textit{False Positives (FP)}, where observations are incorrectly predicted as belonging to the target class; and \textit{False Negatives (FN)}, where observations are incorrectly predicted as not belonging to the target class.
These outcomes are commonly visualized using a confusion matrix. This matrix is created after making predictions on the test data and categorizing each prediction into one of the four possible outcomes mentioned above.

To evaluate model performance, several metrics are commonly used, including accuracy, precision, recall, and the $F_1$ score. Accuracy is defined as the percentage of correct predictions for the test data. It is calculated by dividing the number of correct predictions by the total number of predictions:

\begin{equation}
\text{Accuracy}=\frac{\text{correct predictions}}{\text{all predictions}}
\end{equation}
Precision is defined as the fraction of relevant examples (true positives) among all of the examples that were predicted to belong in a certain class.

\begin{equation}
\text{Precision}=\frac{\text{true positives}}{\text{true positives+false positives}}
\end{equation}
Recall is defined as the fraction of examples that were predicted to belong to a class with respect to all of the examples that truly belong in the class.

\begin{equation}
\text{Recall}=\frac{\text{true positives}}{\text{true positives+false negatives}}
\end{equation}
The $F_1$-score combines a classifier's precision and recall into a single metric by calculating their harmonic mean. It is mainly used to compare the performance of different classifiers. For example, if classifier A has higher recall while classifier B has higher precision, the $F_1$-scores of both classifiers can be compared to determine which one yields better overall results. A score of 1 denotes perfect classification, while lower values indicate poorer classification performance.

\begin{equation}
F_1=\frac{2 \text{ Precision * Recall}}{\text{Precision + Recall}}
\end{equation}
This score is a trustworthy indicator of the overall performance of a classifier.

\subsubsection{Receiver Operating Characteristic (ROC) Curve}

The Receiver Operating Characteristic (ROC) curve is a graphical representation of the diagnostic ability of a binary classifier system as its discrimination threshold is varied. It is a widely used tool in evaluating the performance of classification models, particularly in fields where distinguishing between different classes is critical\citep{Fawcett2006ROC}.
The ROC curve is plotted with the True Positive Rate (TPR) on the y-axis and the False Positive Rate (FPR) on the x-axis. The TPR represents the proportion of correctly identified positive instances out of all actual positives, while the FPR measures the proportion of negative instances incorrectly classified as positive out of all actual negatives \citep{Powers2011ROC}. A perfect classifier would achieve a point at the top-left corner of the plot, corresponding to a TPR of 1 and an FPR of 0.

One of the key metrics derived from the ROC curve is the Area Under the Curve (AUC). The AUC provides a single scalar value that quantifies the overall ability of the classifier to discriminate between the positive and negative classes. An AUC of 0.5 indicates a model with no discriminatory ability, equivalent to random guessing, while an AUC of 1.0 represents a perfect classifier. In practice, classifiers with AUC values closer to 1 are considered better performing \citep{Bradley1997AUC}.
ROC curves are particularly useful when dealing with imbalanced datasets, where the proportion of one class significantly outweighs the other. By focusing on the relative trade-offs between true positives and false positives, ROC analysis provides a comprehensive view of model performance across different classification thresholds \citep{Saito2015AUC}.

In this study, we utilize ROC curves to evaluate the effectiveness of our deep learning model in classifying different cloud types. The AUC values obtained for each classification task are reported in Section \ref{result}, providing insight into the model's discriminative power across the various cloud classes.

\section{Results} \label{result}

\begin{figure}
	\centering
\includegraphics[width=10cm]{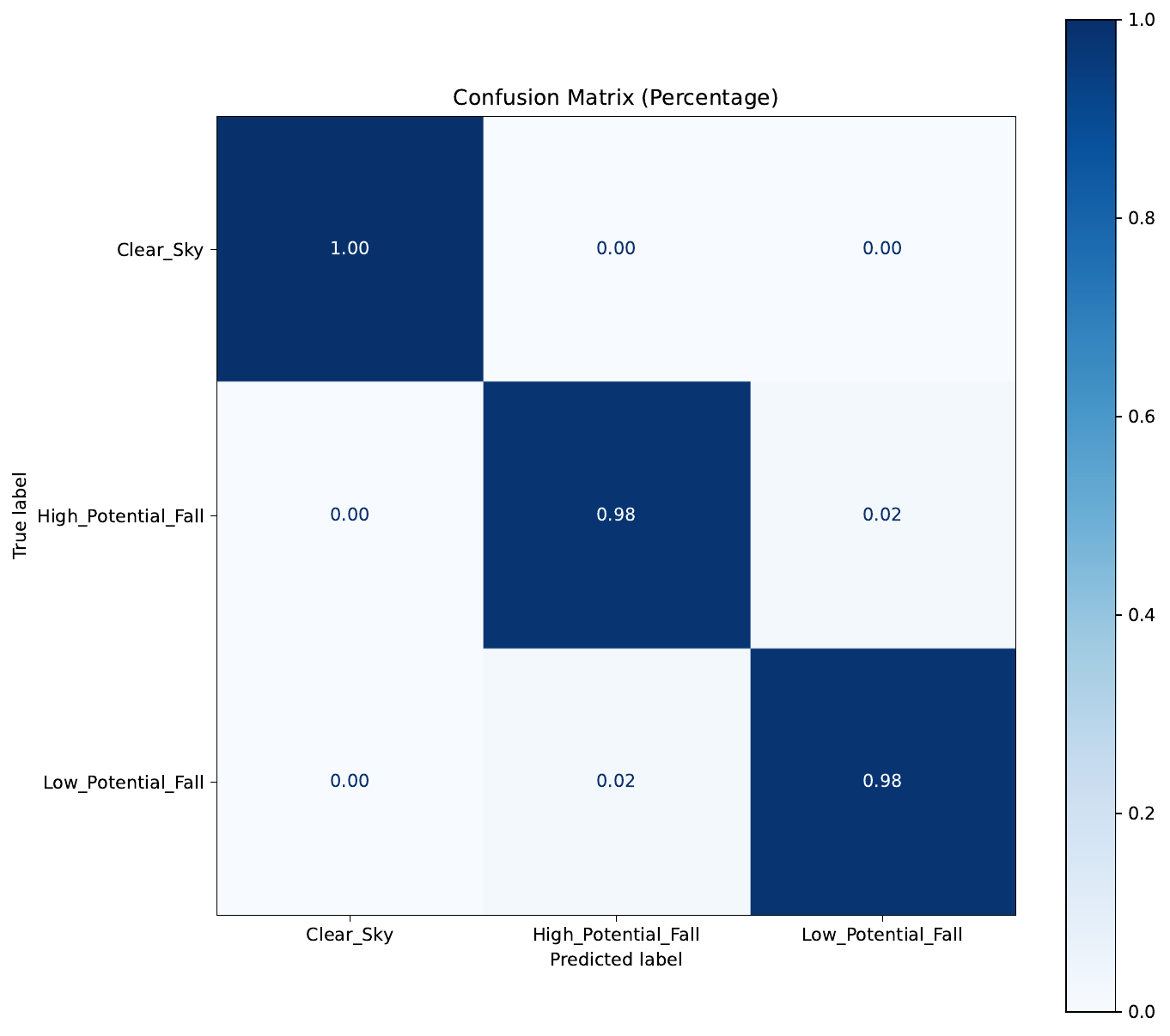}
\caption{Confusion matrix for rainfall.}
\label{cmetrix_rainfall}
\end{figure}

\begin{figure}
	\centering
\includegraphics[width=9cm]{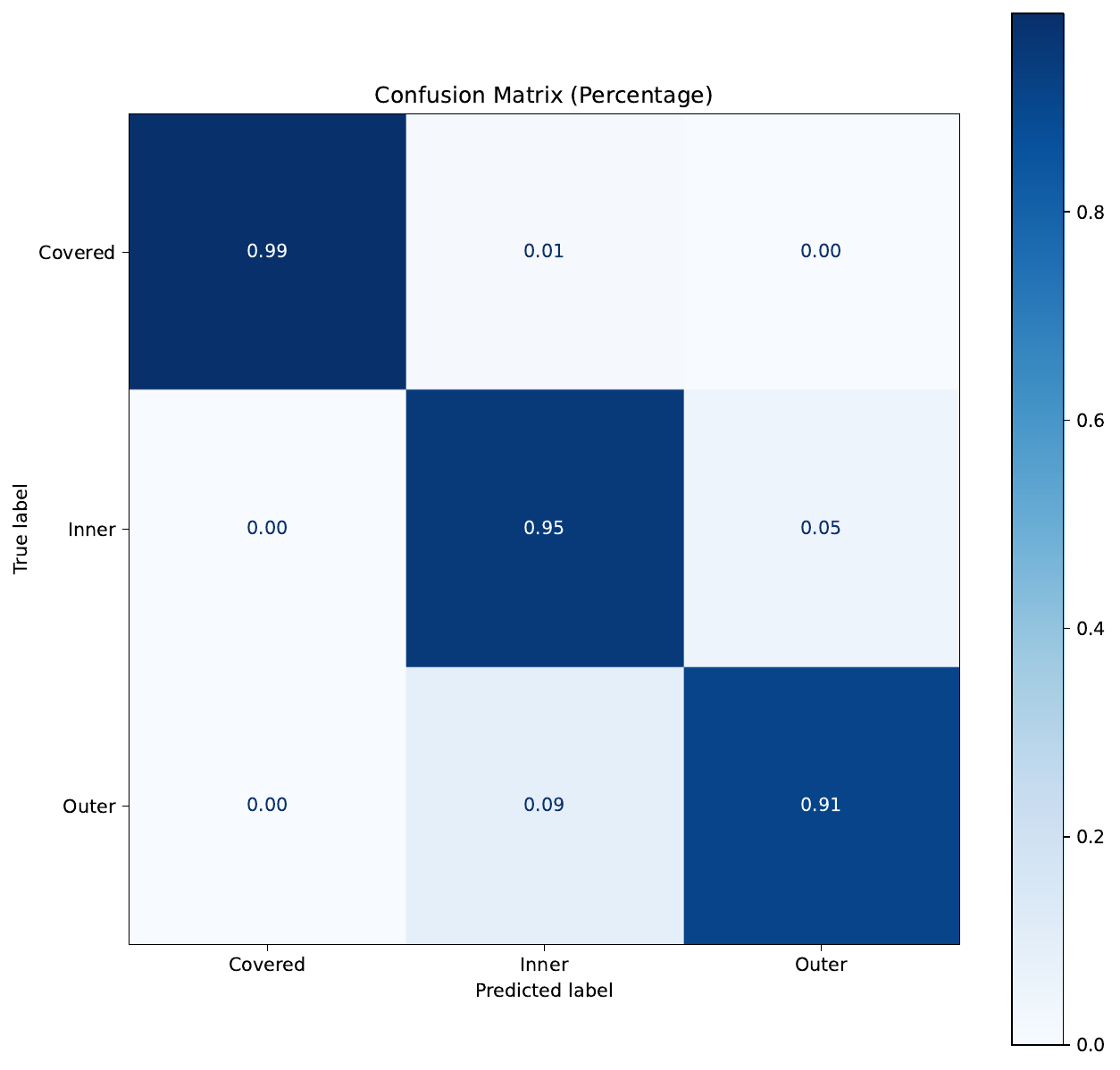}
\caption{Confusion matrix for cloud coverage.}
\label{cmetrix_covrage}
\end{figure}

In the following, we present the results of our model. The test accuracy for cloud coverage classification was 96\%, while the accuracy for rainfall prediction was 99\%. The confusion matrices for both models are shown in Figures \ref{cmetrix_rainfall} and \ref{cmetrix_covrage}, demonstrating how well the model predicts the class of unseen images. For the precipitation model, the Clear\_Sky  class was perfectly distinguished. However, the model made incorrect predictions for 2\% of High\_Potential\_Fall images and 2\% of Low\_Potential\_Fall images.
For the cloud coverage model, there were misclassifications for 1\% of Covered class, 5\% of Inner class and 9\% of Outer class predictions.

\begin{figure}
	\centering
\includegraphics[width=12cm]{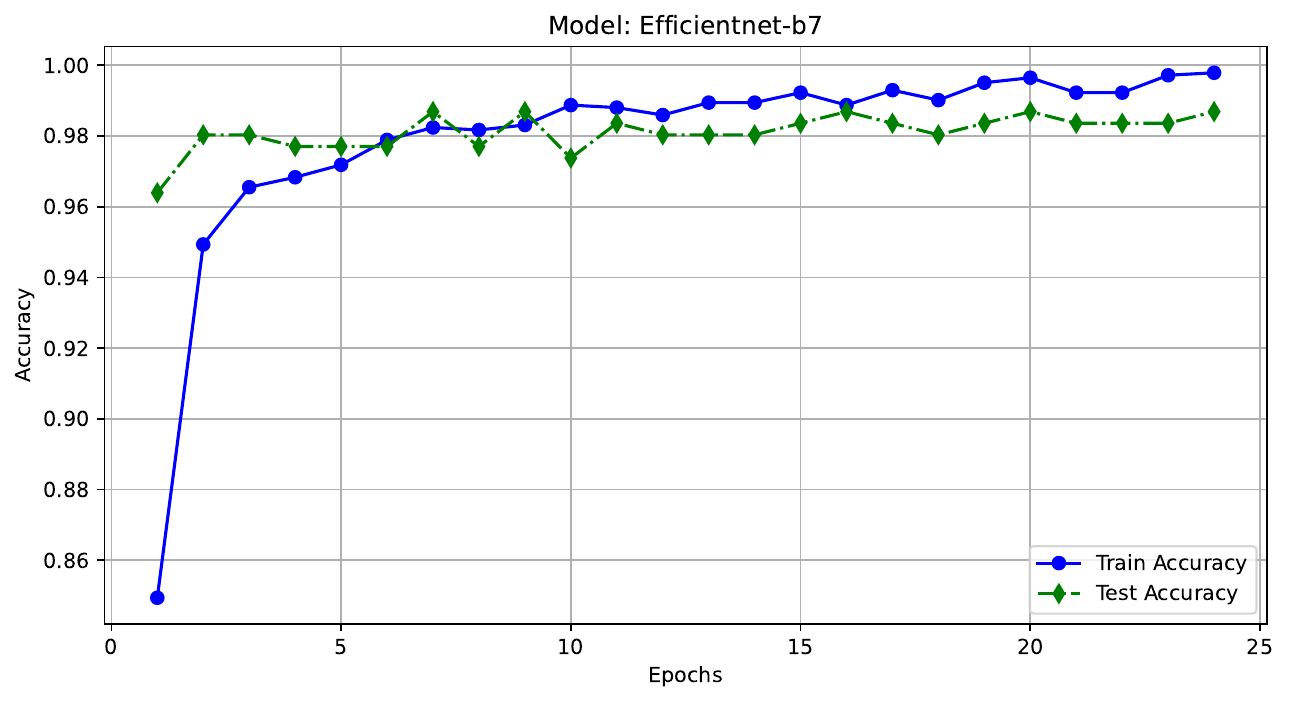}
\caption{Accuracy versus number of epochs for precipitation, which shows the model has converged properly and there is no sign of high bias and variance.}
\label{learn_rainfall}
\end{figure}

\begin{figure}
\centering
\includegraphics[width=12cm]{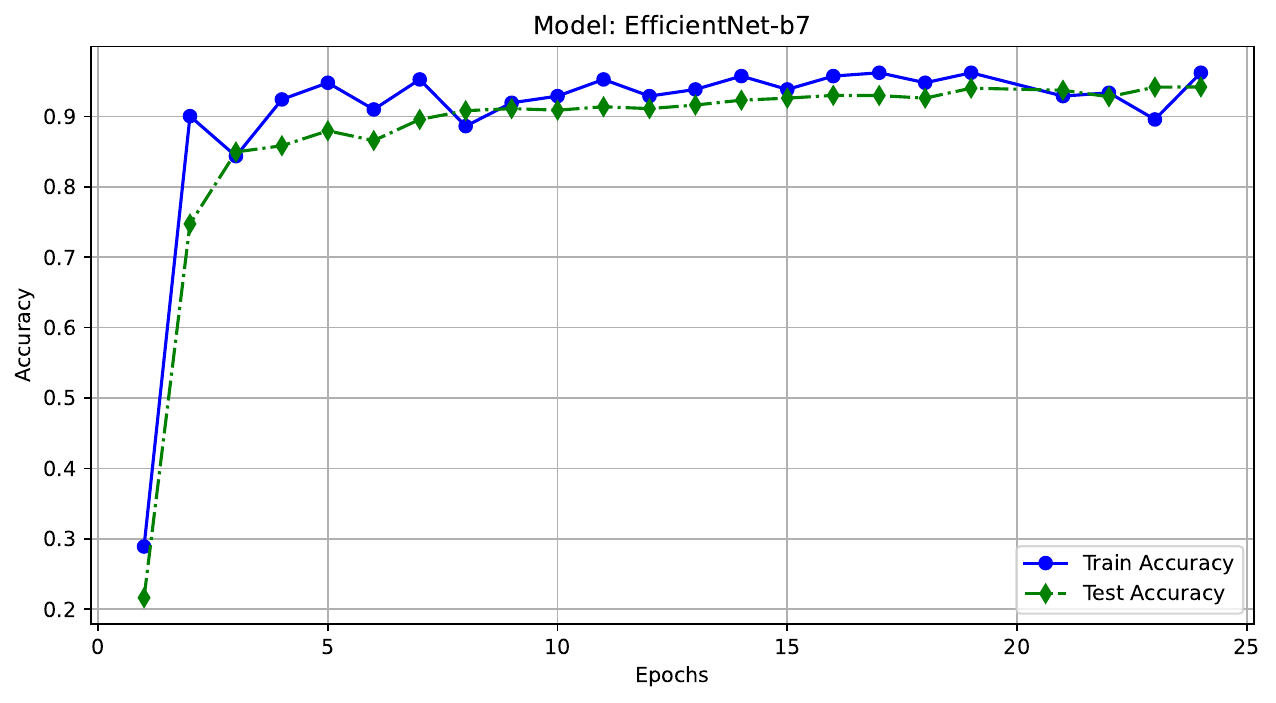}
\caption{Accuracy versus number of epochs for coverage, which shows the model has converged properly and there is no sign of high bias and variance.}
\label{learn_coverage}
\end{figure}

\begin{figure}
	\centering
\includegraphics[width=12cm]{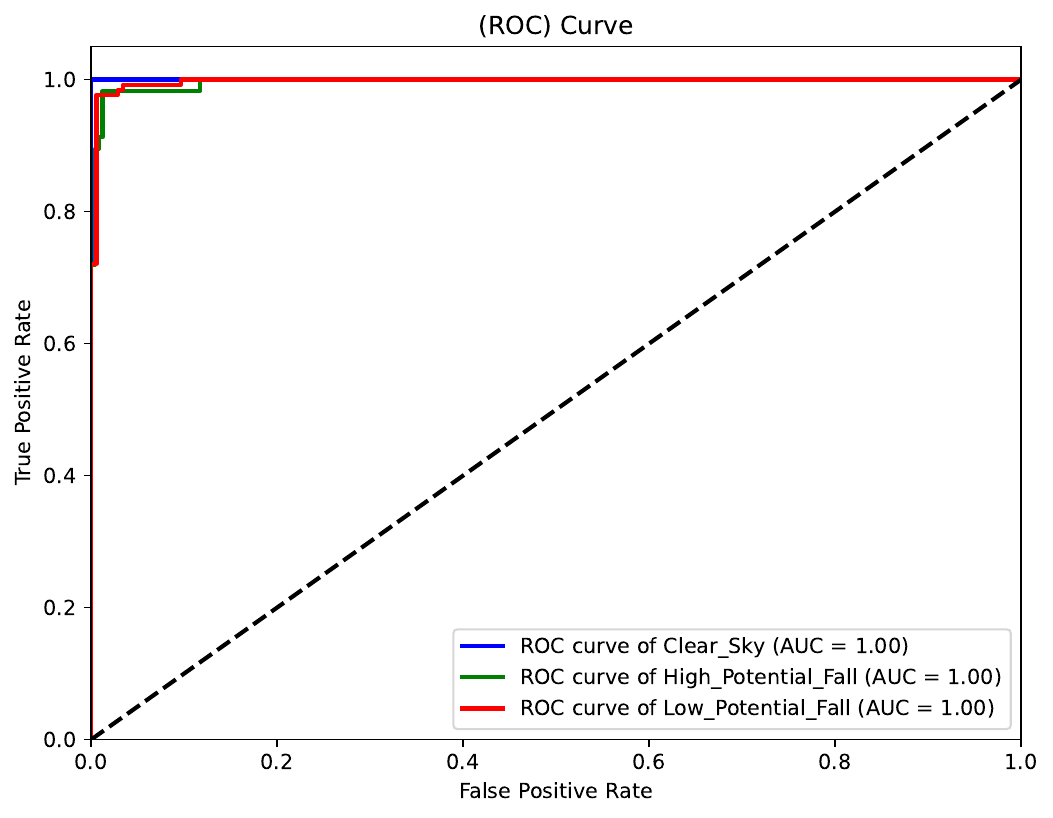}
\caption{Receiver-operating characteristic  (ROC) curve for cloud precipitation.}
\label{roc_rainfall}
\end{figure}

\begin{figure}
	\centering
\includegraphics[width=12cm]{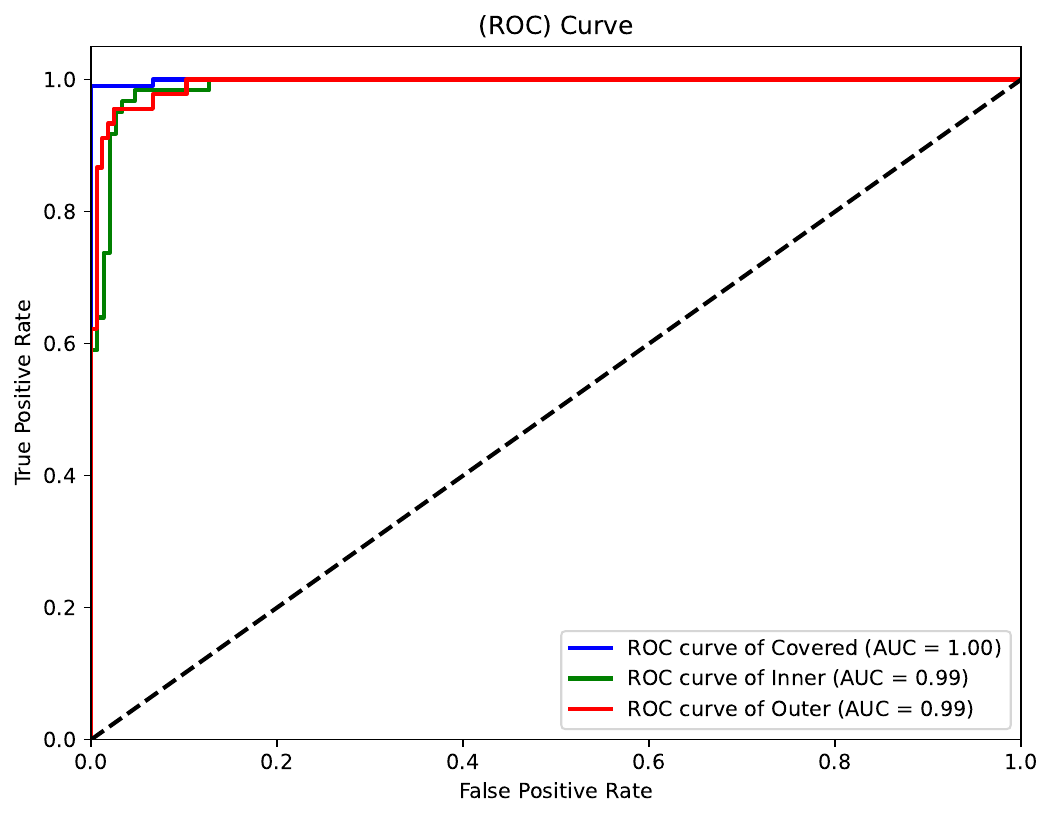}
\caption{Receiver-operating characteristic (ROC) curve for coverage.}
\label{roc_coverage}
\end{figure}

\begin{figure}
	\centering
	\includegraphics[width=14cm]{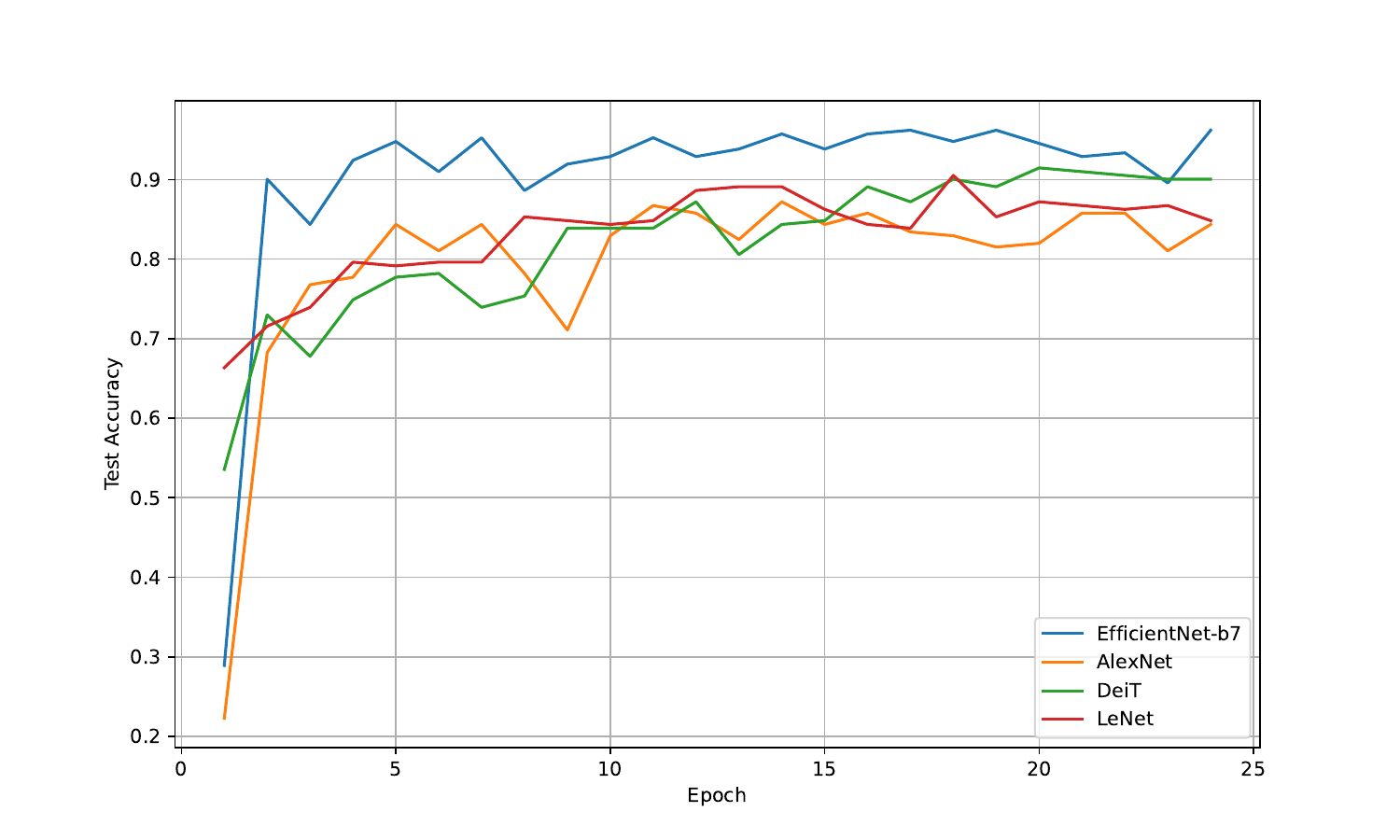}
	\caption{Accuracy of different models versus the number of epochs for cloud coverage classification. }
	\label{models_accu}
\end{figure}

\begin{figure}
	\centering
\includegraphics[width=12cm]{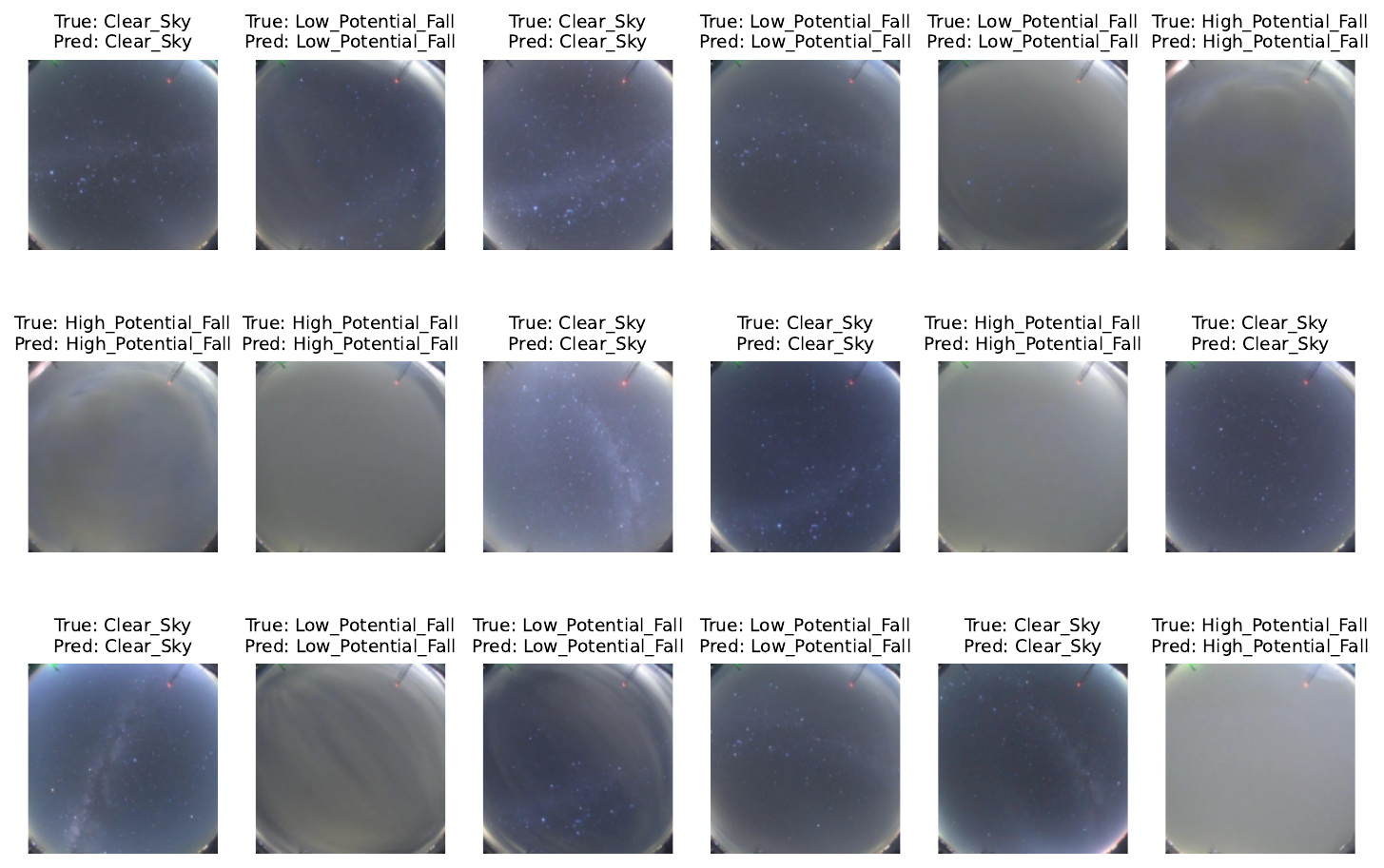}
\caption{Some random images with their corresponding true and predicted precipitation labels.}
\label{True_Pred}
\end{figure}

\begin{figure}
	\centering
\includegraphics[width=12cm]{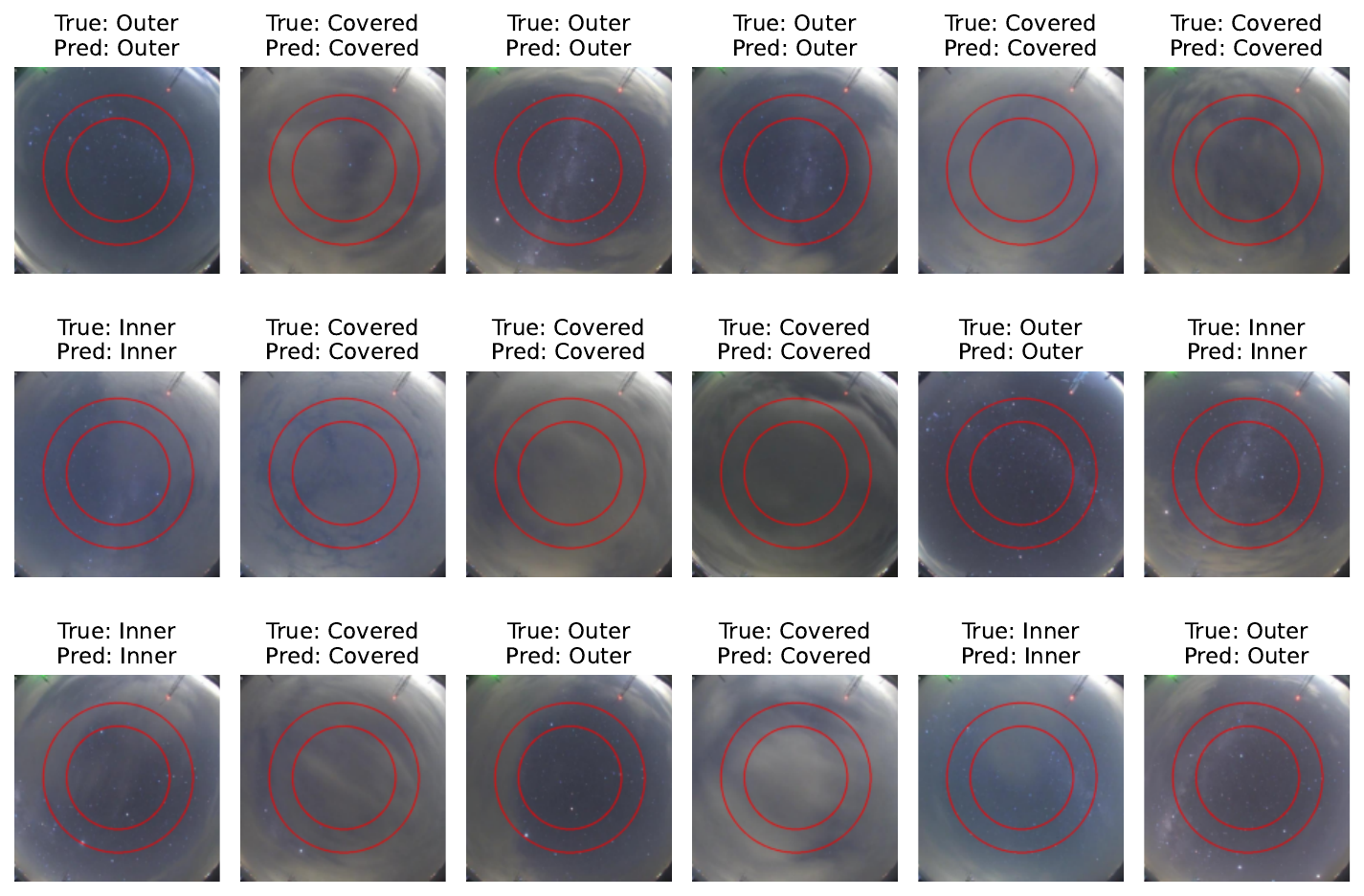}
\caption{Some random images with their corresponding true and predicted coverage labels.}
\label{covrage_circles}
\end{figure}

The learning curves, shown in Figures \ref{learn_rainfall} and \ref{learn_coverage}, indicate that the model has converged with no signs of bias or variance.
Figures \ref{roc_rainfall} and \ref{roc_coverage} display the ROC curves for both tasks. For the first model, the area under the curve (AUC) is  1.00 for all three classes. For the second model, the AUC values are 1.00, 0.99, and 0.99 for the Covered, Inner, and Outer classes, respectively.
The $F_1$ score for precipitation prediction using EfficientNet is 0.98, while for coverage prediction, it is 0.95. To facilitate a comprehensive comparison with other models, we trained several architectures, including LeNet, DeiT, and AlexNet. In tables \ref{tab:model_performance_precipitation}  and \ref{tab:model_performance_coverage}, we present various performance metrics, including accuracy, precision, recall, the  $F_{1}$ score, and the average running time per epoch for all investigated models. The results demonstrate that for precipitation prediction, all models achieve highly satisfactory performance, with the primary distinction being their average running time per epoch. Notably, EfficientNet-b7 exhibits the best performance among the evaluated models. 
In contrast, for the coverage scenario, LeNet, DeiT, and AlexNet show relatively lower success compared to their performance in the precipitation case. Nevertheless, EfficientNet-b7 maintains its superiority, achieving higher performance metrics than the other models. For visual comparison, Figure \ref{models_accu} illustrates the test accuracy of each model as a function of the number of epochs. 
In Figures \ref{True_Pred} \& \ref{covrage_circles}, we provide examples of randomly selected images, along with their predicted and true labels, to visualize the outcomes of the trained models.

\begin{table}[ht]
	\centering
	\caption{Performance metrics for different models for predicting the precipitation.}
	\label{tab:model_performance_precipitation}
	\begin{tabular}{|c|c|c|c|c|c|}
		\hline
		\textbf{Model} & \textbf{Accuracy} & \textbf{Precision} & \textbf{Recall} &  \textbf{F1} & \textbf{Avg Epoch Time } \\ \hline
		EfficinetNet-b7        & 0.987              & 0.986            & 0.980              & 0.983        & 75 s                  \\ \hline
		LeNet        & 0.951               & 0.942            & 0.948              & 0.944        & 9 s                  \\ \hline

		DeiT        & 0.957               & 0.950            & 0.953              & 0.951        & 29 s                  \\ \hline
		AlexNet        & 0.954               & 0.945           & 0.957              & 0.951       & 21 s                  \\ \hline
	\end{tabular}

\end{table}

\begin{table}[ht]
	\centering
	\caption{Performance metrics for different models for predicting the coverage.}
	\label{tab:model_performance_coverage}
	\begin{tabular}{|c|c|c|c|c|c|}
		\hline
	\textbf{Model} & \textbf{Accuracy} & \textbf{Precision} & \textbf{Recall} &  \textbf{F1} & \textbf{Avg Epoch Time } \\ \hline
		EfficinetNet-b7        & 0.962               & 0.951            & 0.950              & 0.951        & 407 s                  \\ \hline
		LeNet        & 0.905               & 0.896            & 0.874             & 0.882        & 16 s                  \\ \hline

		DeiT        & 0.915               & 0.888           & 0.892              & 0.887        & 76 s                  \\ \hline
		AlexNet        & 0.872               & 0.841            & 0.840              & 0.840        & 42 s                  \\ \hline
	\end{tabular}

\end{table}

To further validate the performance of our model and provide additional insight beyond the statistical results, we conducted an analysis of a typical rainfall case using the all-sky image archive. While our primary results are based on a large dataset, this case study offers a qualitative demonstration of the model's ability to predict rainfall events.
We leveraged visual evidence from the all-sky image archive to validate our model's predictions. Specifically, we selected a subset of test images from our dataset for which the model predicted rainfall. We then examined subsequent all-sky images to identify visual imprints of precipitation, such as snowflakes and raindrops or streaks, which typically appear in the minutes or hours following the prediction. Our observations revealed that for the majority of cases where the model predicted precipitation, corresponding imprints were indeed observed in the subsequent images.

\begin{figure*}[htbp]
    \centering
    \begin{tabular}{cc}
        \includegraphics[width=0.35\linewidth]{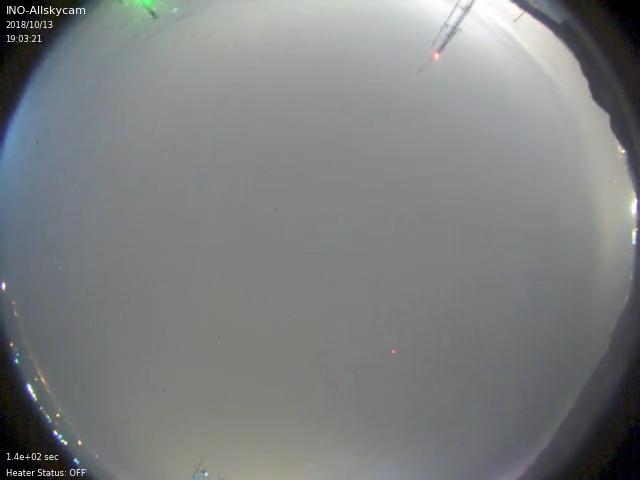} &
        \raisebox{0.5\height}{\vector(1,0){15}}  \textit{Pred: HPF}  \raisebox{0.5\height}{\vector(1,0){15}}  \textit{ After 83 mins }\raisebox{0.5\height}{\vector(1,0){15}}  \includegraphics[width=0.35\linewidth]{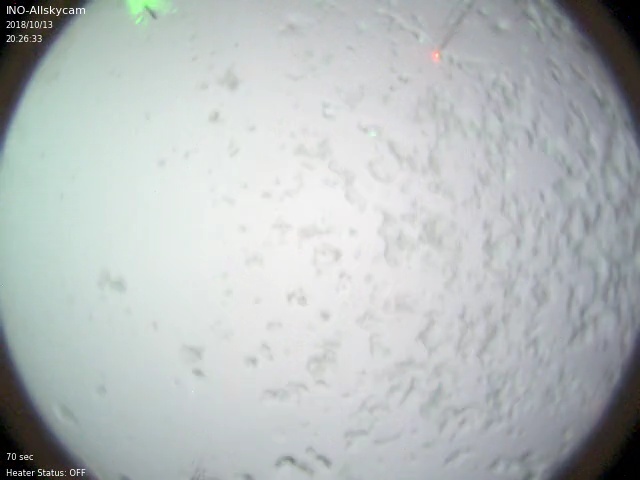} \\
        \includegraphics[width=0.35\linewidth]{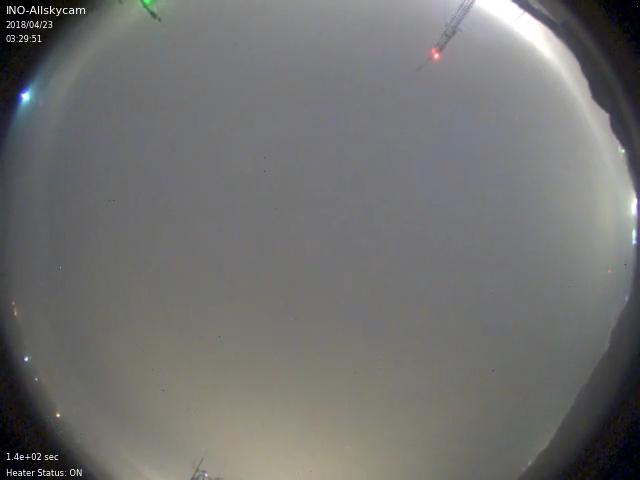} &
        \raisebox{0.5\height}{\vector(1,0){15}}  \textit{Pred: HPF}  \raisebox{0.5\height}{\vector(1,0){15}}  \textit{After 117 mins }\raisebox{0.5\height}{\vector(1,0){15}}  \includegraphics[width=0.35\linewidth]{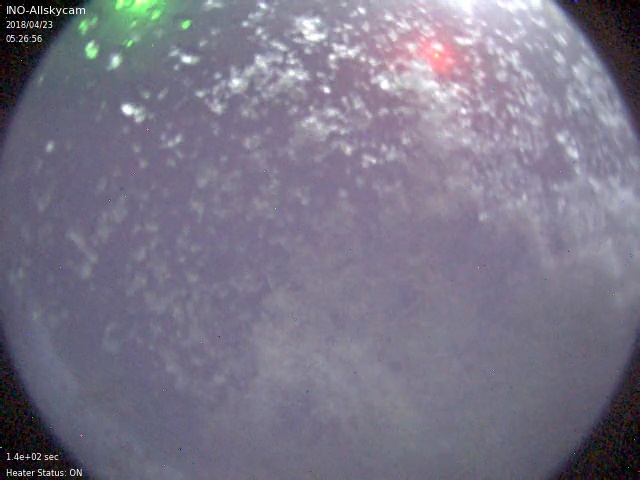} \\
        \includegraphics[width=0.35\linewidth]{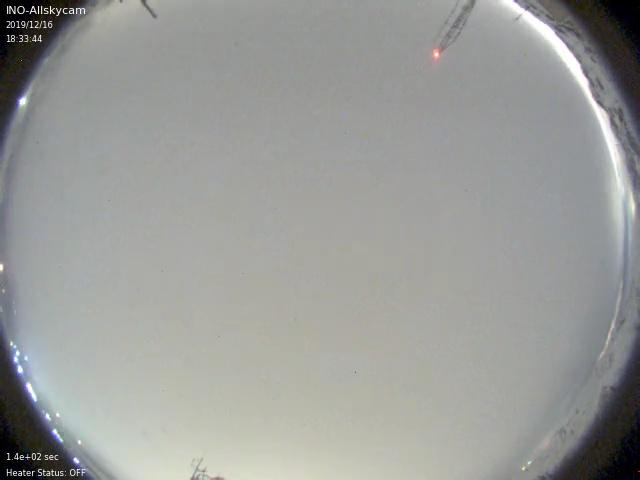} &
        \raisebox{0.5\height}{\vector(1,0){15}}  \textit{Pred: HPF}  \raisebox{0.5\height}{\vector(1,0){15}}  \textit{After 208 mins }\raisebox{0.5\height}{\vector(1,0){15}}  \includegraphics[width=0.35\linewidth]{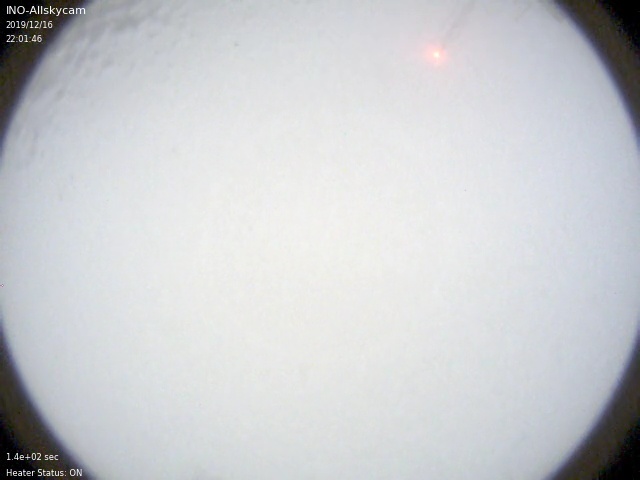} \\
        \includegraphics[width=0.35\linewidth]{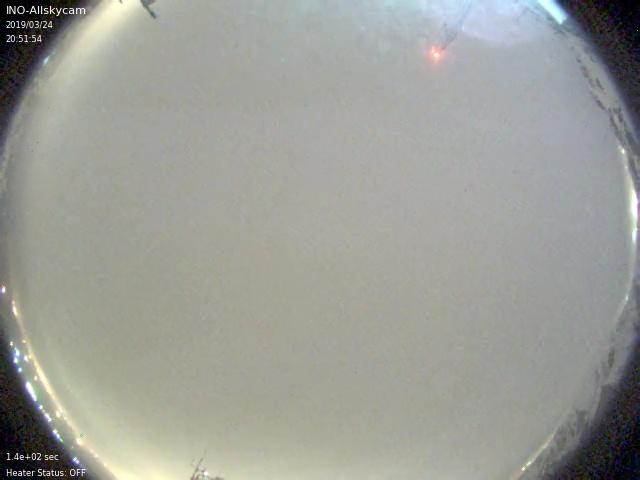} &
        \raisebox{0.5\height}{\vector(1,0){15}}  \textit{Pred: HPF}  \raisebox{0.5\height}{\vector(1,0){15}}  \textit{After 212 mins }\raisebox{0.5\height}{\vector(1,0){15}}  \includegraphics[width=0.35\linewidth]{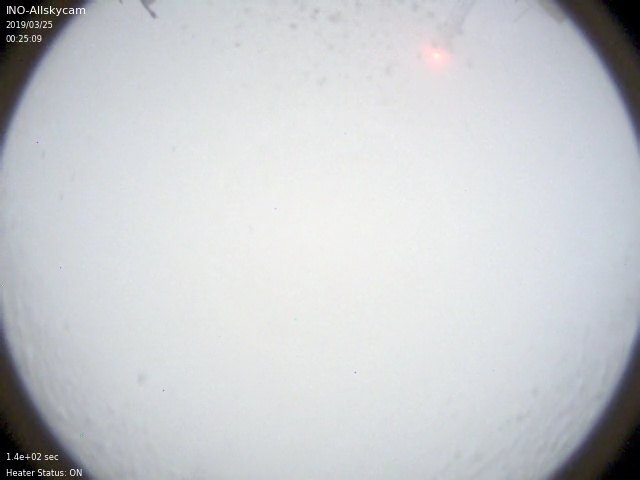} \\
    \end{tabular}
    \caption{Visual demonstration of the model's precipitation predictions. The left column displays all-sky images used for making predictions, while the right column shows subsequent images with visible precipitation imprints. The label "HPF" refers to High Probability Rainfall prediction. The time interval between the model's prediction and the observed precipitation imprints is also provided.}
    \label{fig:rainfall_predictions}
\end{figure*}

To illustrate this, Figure~\ref{fig:rainfall_predictions} shows an example of the model's prediction alongside subsequent all-sky images with visible rainfall and snowfall imprints. The left panels display the all-sky images for which the model predicted precipitation, while the right panels show images captured in the following minutes, clearly depicting precipitation imprints. This analysis provides a practical demonstration of the model's capability to predict precipitation events based on all-sky camera images and its potential for real-world applications.

\section{Summary and Conclusions}

In this manuscript, we present an automated method for identifying  precipitation clouds using deep learning algorithms. This approach offers an accurate and cost-effective solution for protecting and maintaining telescopes and observational instruments at ground-based observatories. As demonstrated in this study, via EfficientNet-b7, precipitation potential can be predicted with 99\% accuracy and cloud coverage with 96\% accuracy, enabling prompt closure and reopening of telescope domes. For the sake of comparison and to explore the performance of alternative architectures on the task, we trained three additional models: LeNet, DeiT, and AlexNet. For precipitation prediction, all models performed well, with EfficientNet-b7 emerging as the top performer. However, in terms of coverage, LeNet, DeiT, and AlexNet achieved $F_{1}< 0.9$ while  EfficientNet-b7 achieved $F_{1}= 0.95$.
The method developed in this study can improve telescope efficiency and maximize available observation time. Using this approach, three scenarios are possible:

First, if a cloud is predicted as High\_Potential\_Fall, the dome can be automatically closed, regardless of whether the cloud is classified as Inner, Outer, or Covered. Second, if a cloud is predicted as Low\_Potential\_Fall, depending on whether its coverage is classified as Outer or Inner, observations can either continue or be halted. Finally, if an all-sky image is predicted as Clear\_Sky, observations can proceed uninterrupted. Since no specific image pre-processing was performed on the training set and the models classify images quickly, this approach can be used for real-time applications.

With a large number of space and ground-based astronomical instrumentation publishing alerts, most small to medium-class observatories are adopting robotic or automated operation modes in which manual human intervention is reduced to supervise the entire operation. Automated, reliable, and robust cloud detection methods that can maximize observing time while safeguarding the equipment are becoming increasingly important.

\section*{Acknowledgement}
We would like to thank the Iranian National Observatory for data access. We would to extend our thanks to Hamed Altafi and Alireza Behnam.

{}


\begin{thebibliography}{}
	
	\bibitem[Afiq et al.(2019)]{afiq2019} Afiq, M., Hamid, A., Mohd, W., \& Mohamad, N. S. (2019). Urban night sky conditions determination method based on a low-resolution all-sky images. \textit{IEEE Conference Proceedings}, 158–162.
	
	\bibitem[Bradley(1997)]{Bradley1997AUC} Bradley, A. P. (1997). The use of the area under the ROC curve in the evaluation of machine learning algorithms. \textit{Pattern Recognition, 30}(7), 1145–1159. https://doi.org/10.1016/S0031-3203(96)00142-2
	
	\bibitem[Calbó \& Sabburg(2008)]{calbo} Calbó, J., \& Sabburg, J. (2008). Feature detection in all-sky images. \textit{Journal of Atmospheric and Oceanic Technology, 25}(1), 3–17. https://doi.org/10.1175/2007JTECHA959.1
	
	\bibitem[Chen et al.(2020)]{chen2020efficientnet} Chen, C., Dou, Q., Chen, H., Qin, J., \& Heng, P.-A. (2020). EfficientNet: Rethinking model scaling for convolutional neural networks. \textit{Medical Image Analysis, 61}, 101665. https://doi.org/10.1016/j.media.2020.101665
	
	\bibitem[Dev et al.(2017)]{dev2017} Dev, S., Savoy, F. M., Lee, Y.-H., \& Winkler, S. (2017). Nighttime cloud detection with all-sky images using a convolutional neural network. In \textit{Proceedings of the IEEE International Conference on Image Processing (ICIP)} (pp. 345–349). https://doi.org/10.1109/ICIP.2017.8296257
	
	\bibitem[Dosovitskiy et al.(2020)]{dosovitskiy2020image} Dosovitskiy, A., Beyer, L., Kolesnikov, A., Weissenborn, D., Zhai, X., Unterthiner, T., Dehghani, M., Minderer, M., Heigold, G., Gelly, S., \& Houlsby, N. (2020). An image is worth 16x16 words: Transformers for image recognition at scale. \textit{arXiv preprint arXiv:2010.11929}.
		
		
	\bibitem[Fawcett(2006)]{Fawcett2006ROC} Fawcett, T. (2006). An introduction to ROC analysis. \textit{Pattern Recognition Letters, 27}(8), 861–874. https://doi.org/10.1016/j.patrec.2005.10.010
	
	\bibitem[Goodfellow et al.(2016)]{Goodfellow} Goodfellow, I., Bengio, Y., \& Courville, A. (2016). \textit{Deep Learning}. MIT Press.
	
	\bibitem[Guzel et al.(2024)]{Guzel} Guzel, M., Kalkan, M., Bostanci, E., Acici, K., \& Asuroglu, T. (2024). Advances in machine learning for cloud classification. \textit{PeerJ Computer Science, 10}, e1779. https://doi.org/10.7717/peerj-cs.1779
	
	\bibitem[Hu et al.(2018)]{hu2018squeeze} Hu, J., Shen, L., \& Sun, G. (2018). Squeeze-and-excitation networks. \textit{IEEE Transactions on Pattern Analysis and Machine Intelligence, 42}(8), 2011–2023. https://doi.org/10.1109/TPAMI.2019.2913372
	
	\bibitem[Jadhav \& Aditi(2015)]{jadhav2015} Jadhav, T., \& Aditi, K. (2015). Cloud classification in all-sky images. In \textit{Proceedings of the 8th International Conference on Contemporary Computing (IC3)} (pp. 275–278). Noida, India.
	
	\bibitem[Kingma \& Ba(2014)]{Kingma2014Adam} Kingma, D. P., \& Ba, J. L. (2014). Adam: A method for stochastic optimization. \textit{arXiv Preprint}. https://doi.org/10.48550/arXiv.1412.6980
	
\bibitem[Khosroshahi et al.(2022)]{Khosroshahi} 
Khosroshahi, H. G., Bidar, M., Jenab, H., Ravanmehr, R., Mohajer, M., Saeidifar, M., Behnam, A., \& Shomali, R. (2022). Iranian National Observatory: project overview and achievements. In H. K. Marshall, J. Spyromilio, \& T. Usuda (Eds.), \textit{Ground-based and Airborne Telescopes IX} (Vol. 12182, p. 121820Y). Society of Photo-Optical Instrumentation Engineers (SPIE) Conference Series. doi:10.1117/12.2628800
	
	
	\bibitem[Krizhevsky et al.(2012)]{krizhevsky2012imagenet} Krizhevsky, A., Sutskever, I., \& Hinton, G. E. (2012). ImageNet classification with deep convolutional neural networks. In \textit{Advances in Neural Information Processing Systems (Vol. 25)}, 1097–1105.
	
	\bibitem[LeCun et al.(1998)]{lecun1998gradient} LeCun, Y., Bottou, L., Bengio, Y., \& Haffner, P. (1998). Gradient-based learning applied to document recognition. \textit{Proceedings of the IEEE, 86}(11), 2278–2324.
	
	
	\bibitem[Li et al.(2022)]{li2022} Li, X. T., Wang, B. Z., Qiu, B., \& Wu, C. (2022). Advances in all-sky cloud detection techniques. \textit{Atmospheric Measurement Techniques, 15}, 3629–3640. https://doi.org/10.5194/amt-15-3629-2022
	
	\bibitem[Markowski \& Richardson(2010)]{markowski2010} Markowski, P., \& Richardson, Y. (2010). \textit{Mesoscale Meteorology in Midlatitudes}. Wiley-Blackwell.
	
	\bibitem[Mommert(2020)]{momert} Mommert, M. (2020). Advances in all-sky imaging techniques. \textit{Astronomical Journal, 159}(4), 178. https://doi.org/10.3847/1538-3881/ab744f
	
	\bibitem[Powers(2011)]{Powers2011ROC} Powers, D. M. (2011). Evaluation: From precision, recall and F-measure to ROC, informedness, markedness, and correlation. \textit{Journal of Machine Learning Technologies, 2}(1), 37–63.
	
	\bibitem[Ramachandran et al.(2017)]{ramachandran2017swish} Ramachandran, P., Zoph, B., \& Le, Q. V. (2017). Swish: A self-gated activation function. \textit{arXiv Preprint}. https://doi.org/10.48550/arXiv.1710.05941
	
	\bibitem[Rogers \& Yau(2000)]{rogers2000} Rogers, R. R., \& Yau, M. K. (2000). \textit{A Short Course in Cloud Physics} (3rd ed.). Elsevier.
	
	\bibitem[Saito \& Rehmsmeier(2015)]{Saito2015AUC} Saito, T., \& Rehmsmeier, M. (2015). The precision-recall plot is more informative than the ROC plot when evaluating binary classifiers on imbalanced datasets. \textit{PLOS ONE, 10}(3), e0118432. https://doi.org/10.1371/journal.pone.0118432
	
	\bibitem[Shi et al.(2021)]{shi2021} Shi, C. J., Zhou, Y. T., Qiu, B., Guo, D. J., \& Li, M. C. (2021). Enhanced cloud detection methods. \textit{IEEE Geoscience and Remote Sensing Letters, 18}(11), 1688–1691. https://doi.org/10.1109/LGRS.2020.3010409
	
	\bibitem[Singh \& Glennen(2005)]{Glennen} Singh, M., \& Glennen, M. (2005). Image segmentation with deep neural networks. \textit{Pattern Analysis and Applications, 8}(3), 258–271. https://doi.org/10.1007/s10044-005-0007-5
	
	\bibitem[Stull(2015)]{stull2015} Stull, R. B. (2015). \textit{Practical Meteorology: An Algebra-based Survey of Atmospheric Science}. University of British Columbia.
	
	\bibitem[Tan \& Le(2019)]{efficientnet} Tan, M., \& Le, Q. V. (2019). EfficientNet: Rethinking model scaling for convolutional neural networks. In \textit{Proceedings of the 36th International Conference on Machine Learning (ICML)} (pp. 6105). https://doi.org/10.48550/arXiv.1905.11946
	
	\bibitem[Touvron et al.(2021)]{touvron2021training} Touvron, H., Cord, M., Douze, M., Massa, F., Sablayrolles, A., \& Jégou, H. (2021). Training data-efficient image transformers \& distillation through attention. In \textit{Proceedings of the International Conference on Machine Learning}, 10347–10357.
		
	\bibitem[Wallace \& Hobbs(2006)]{wallace2006} Wallace, J. M., \& Hobbs, P. V. (2006). \textit{Atmospheric Science: An Introductory Survey} (2nd ed.). Elsevier.
	
	\bibitem[WMO(1987)]{WMO} WMO. (1987). \textit{International Cloud Atlas, Vol. 2}. American Meteorological Society.
	
	\bibitem[Yosinski et al.(2014)]{Yosinski} Yosinski, J., Clune, J., Bengio, Y., \& Lipson, H. (2014). How transferable are features in deep neural networks? \textit{arXiv Preprint}. https://doi.org/10.48550/arXiv.1411.1792
	
	\bibitem[Yousaf et al.(2023)]{Yousaf} Yousaf, R., Rehman, H. Z. U., Khan, K., et al. (2023).
	
	

	


	

	
	
\end{thebibliography}
\end{document}